\documentclass[pra,10pt,twocolumn,groupedaddress,floatfix,showpacs]{revtex4-1}

\usepackage[utf8]{inputenc}  
\usepackage[sc,osf]{mathpazo}
\usepackage[scaled=0.86]{berasans}  
\usepackage[colorlinks=true, allcolors=blue, urlcolor=blue]{hyperref}  
\usepackage{graphicx} 
\usepackage[babel]{microtype}  
\usepackage{amsmath,amssymb,amsthm,bm,amsfonts,mathrsfs,bbm} 

\usepackage{xspace}  
\usepackage{pgfplots}
\usepackage{xcolor,colortbl}
\usepackage{array}
\usepackage{bigstrut}
\usepackage{tcolorbox}

\newcommand{\tr}{\text{tr}}
\newcommand{\ket}[1]{| #1 \rangle}

\newcommand{\be}{\begin{equation}}
\newcommand{\ee}{\end{equation}}
\newcommand{\bea}{\begin{eqnarray}}
\newcommand{\eea}{\end{eqnarray}}
\newcommand{\bes}{\begin{equation*}}
\newcommand{\ees}{\end{equation*}}
\newcommand{\beas}{\begin{eqnarray*}}
	\newcommand{\eeas}{\end{eqnarray*}}

\usepackage{amsmath}
\usepackage{mathtools}
\usepackage{amssymb}

\newcommand{\x}{\mathrm{x}}

\def\x{\mathrm{x}}

\def\tr{\mathrm{tr}}

\newcommand{\qgate}[4]{\begin{pmatrix} #1 & #2 \\ #3 & #4 \end{pmatrix}}

\newtheorem*{thm*}{Theorem}

\newtheorem*{lem*}{Lemma}

\newtheorem*{lipschitzLem*}{Lemma \ref{lipschitz}}
\newtheorem*{lipschitzCubeLem*}{Lemma \ref{lipschitzCube}}
\newtheorem*{pgmNearlyOptimalThm*}{Theorem \ref{pgmNearlyOptimal}}

\begin{document}

\title{ Detecting Entanglement Generating Circuits in Cloud-Based Quantum Computing }


\author{Jiheon Seong}
\email{jiheon94@kaist.ac.kr}

\author{Joonwoo Bae}
\email{joonwoo.bae@kaist.ac.kr}

\affiliation{School of Electrical Engineering, Korea Advanced Institute of Science and Technology (KAIST), 291 Daehak-ro, Yuseong-gu, Daejeon 34141, Republic of Korea }


\begin{abstract}
Entanglement, a direct consequence of elementary quantum gates such as controlled-NOT or Toffoli gates, is a key resource that leads to quantum advantages. In this work, we establish the framework of certifying entanglement generation in cloud-based quantum computing services. Namely, we present the construction of quantum circuits that certify entanglement generation in a circuit-based quantum computing model. The framework relaxes the assumption of the so-called qubit allocation, which is the step in a cloud service to relate physical qubits in hardware to a circuit proposed by a user. Consequently, the certification is valid no matter how unsuccessful qubit allocations may be in cloud computing or how untrustful the service may be in qubit allocations. We then demonstrate the certification of entanglement generation on two and three qubits in the IBMQ and IonQ services. Remarkably, entanglement generation is successfully certified in the IonQ service that does not provide a command of qubit allocations. The capabilities of entanglement generation in the circuits of IBMQ and IonQ are also quantified. We envisage that the proposed framework is applied when cloud-based quantum computing services are exploited for practical computation and information tasks, for which our results would find if it is possible to achieve quantum advantages. 
\end{abstract}


\maketitle

\section{Introduction}

Efficient information processing beyond the existing limitations may be achieved by applying the laws of quantum mechanics as the working principles \cite{Deutsch:1985wz}. For instance, the prime-number factorization problem, considered to be hard with conventional computers, turns out to be tractable in a quantum computer \cite{Shor:1997ul}. Quantum database search is more efficient than its classical counterpart by showing a quadratic speedup \cite{PhysRevLett.79.325}.  

The currently available quantum technologies, often referred to as noisy intermediate scale quantum (NISQ) technologies \cite{Preskill2018quantumcomputingin}, yet contain noise in all of the steps, the preparation of qubits, a quantum circuit, and a measurement. It has been shown that, remarkably, NISQ systems can work beyond the classical limitations by having the laws of quantum mechanics as the working principles in computational tasks. It is immediately possible to achieve quantum advantages over the classical counterparts \cite{Arute:2019ts, Harrow:2017ve, Bravyi308}. Prototype quantum computers based on the NISQ technologies are also available from industry vendors, such as IBMQ \cite{ibmq} and IonQ \cite{ionq}. They can be accessed in the form of cloud-based quantum computing. 

It appears that in fact, entanglement has been identified as a general and key resource that leads to the quantum advantages \cite{RevModPhys.81.865}. Since entangled states are a direct consequence of universal quantum gates such as controlled-NOT or Toffoli gates \cite{AHARONOV:1999um}, the presence of entanglement also verifies the capabilities of the universal quantum gates. Entangled states can be used to enhance the performance of various quantum information applications, such as quantum channels, networks, quantum sensing, and quantum computing \cite{PhysRevLett.122.140404, PhysRevLett.122.140402, PhysRevLett.122.140403}. Arbitrarily weakly entangled states are also useful \cite{PhysRevLett.86.544}. Once entangled states are detected, they can be manipulated and distilled into maximally entangled states \cite{PhysRevLett.76.722}. 

Recently, quantum algorithms fitted to NISQ technologies have been proposed, namely, hybrid quantum-classical algorithms \cite{bharti2021noisy} such as quantum approximate optimization algorithms (QAOA) \cite{farhi2014quantum} and variational quantum eigensolver (VQE) \cite{Peruzzo:2014uz}: shallow quantum circuits that are well-designed within the capability of NISQ technologies are repeatedly applied. One can find that the building blocks are characterized by parameterized quantum circuits (PQCs) composed of single- and two-qubit gates with a polynomial number of control parameters. it has been shown that PQCs can be used in practical quantum computing applications, e.g., machine learning, neural networks, and optimization \cite{Benedetti:2019tu, PhysRevA.98.062324, PhysRevX.7.041052, PhysRevLett.122.040504, Huggins:2019wc, Grant:2018vv}. The capability of generating entangled states is one of the  key properties that enable PQCs to lead to quantum advantages \cite{https://doi.org/10.1002/qute.201900070, PhysRevResearch.2.033125}.

\begin{figure*}[t]
	\begin{center}
		\includegraphics[angle=0, width=1.\textwidth]{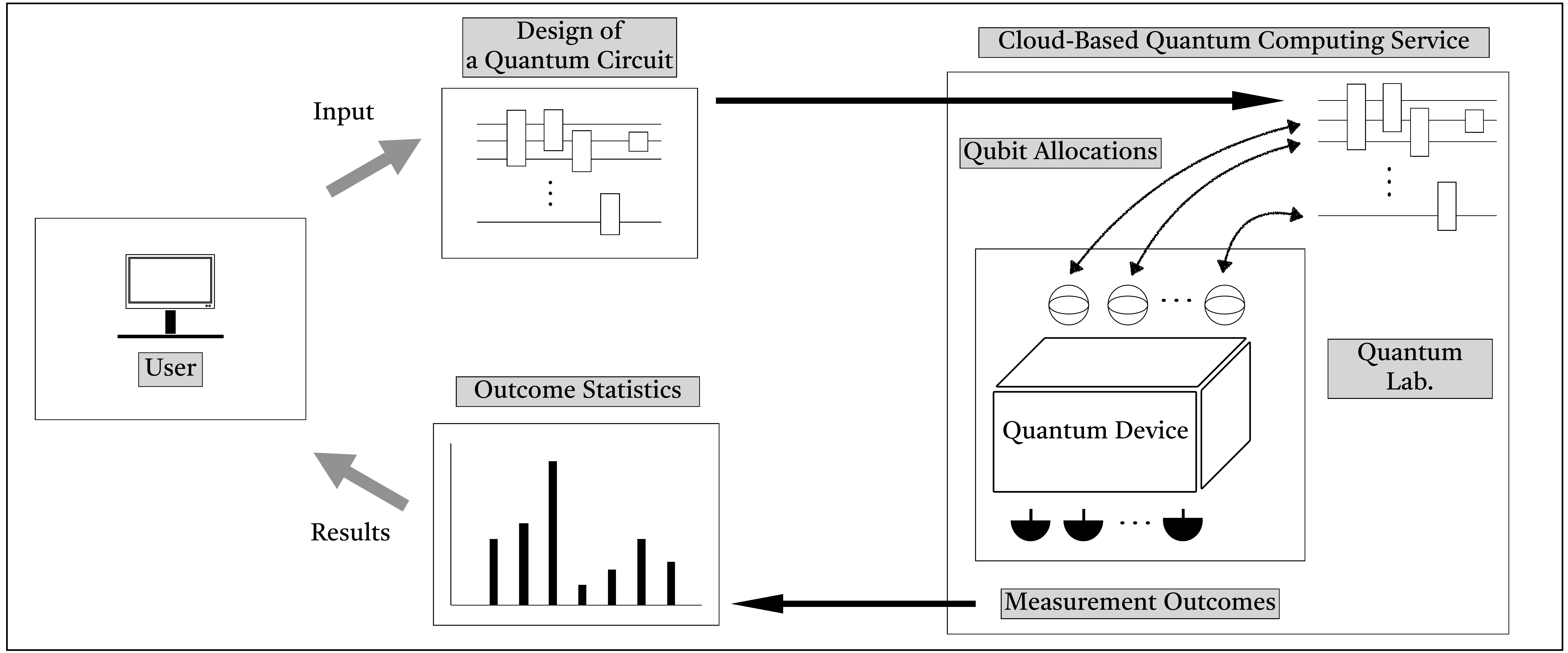}
		\caption{The scenario of a cloud-based quantum computing service is summarized. The input to the service by a user is a design of a quantum circuit. The service begins with the step called qubit allocations that relate physical qubits to a designed circuit. An implementation of the circuit is run in the service, and then measurement statistics returns to the user as the outcome. Given the two types of data, a computational task is concluded. Quantum advantages may be obtained when quantum circuits capable of generating entangled states are realized in the service. This work presents the architecture of quantum circuits for certifying entanglement generation in cloud-based quantum computing services. } \label{sce}
	\end{center}	
\end{figure*} 

All these signify the importance of certifying entanglement generation in practice, in particular, in the prototype quantum computers with NISQ technologies. The certification of entanglement generation in quantum hardware shows a realization of two- and three-qubit elementary gates. It also implies the practical usefulness of quantum hardware at the most fundamental level toward quantum advantages in real applications. 

A straightforward strategy to find entanglement generation by a circuit-based quantum computing model would be to perform quantum process tomography that identifies an experimental quantum operation acted on qubits \cite{PhysRevLett.78.390, estimation}. Once a quantum operation is fully characterized, one may find if the verified dynamics can generate entangled states. However, the strategy is neither practical nor efficient due to the required experimental resources. Quantum process tomography of $n$-qubit dynamics requires $O(2^{2n})$ measurements as well as an inverse problem involved in the estimation.  

One may consider entanglement witnesses (EWs) in a quantum circuit and realize them in a cloud-based quantum computing service, see e.g. EWs in the IBMQ service \cite{Mooney:2019tt}. We recall that EWs are a well-established tool to detect entangled states without the full knowledge via quantum tomography \cite{TERHAL2000319, GUHNE20091, MR3280004}. 

\subsection*{Qubit allocations}

Note, however, that a cloud computing service does not precisely correspond to a standard scenario of a quantum laboratory. This is because of an additional step called a command of qubit allocations, which relate a cloud computing service to physical qubits in hardware according to a circuit sent by a user.

{\it The protocol} of the cloud-based quantum computing services is summarized as follows. 
\begin{itemize}
\item Step1. A user prepares a design of a quantum circuit from a mathematical problem of interest. 
\item Step2. A cloud-based quantum computing service runs an implementation of the circuit in its devices and collects the outcome statistics.  
\item Step3. A user acknowledges the statistics of measurement outcomes.
\end{itemize}
As for a user in a cloud service, {\it a design of a quantum circuit and the outcome statistics} only are given, in which the mapping between a quantum circuit and quantum hardware is performed by the service via the command of qubit allocations. 

\begin{tcolorbox}
Qubit allocations: A cloud-based quantum computing service relates physical qubits in hardware to a circuit proposed by a user. 
\end{tcolorbox}

Then, in a realization EWs in cloud-based quantum computing service, we point out that it is necessary to trust qubit allocations. The reason is that, since an EW is not positive, its expectation value can be estimated by different measurements: in each measurement the same qubits should be addressed. To make the point clear, let us consider a realization of EWs, as follows.

Suppose that it is aimed to find entanglement generation among qubits labeled by $A$ and $B$. Technically, an EW corresponds to a Hermitian operator $W^{AB}$ such that 
\begin{eqnarray}
\mathrm{tr} [W^{AB} \sigma_{\mathrm{sep}}^{AB} ] \geq 0 ~\mathrm{for~all~separable~states~}\sigma_{\mathrm{sep}}^{AB}. \label{eq:1}
\end{eqnarray}
The task is to estimate an expectation value, $\tr[W^{AB}\rho^{AB}]$, for an unknown state $\rho^{AB}$: negative values certify that the state $\rho^{AB}$ is entangled. As an EW is not positive, it can be implemented via its decomposition into positive and negative projections, i.e., 
\bea
W^{^{AB}} = c_+P_{+}^{AB} - c_-P_{-}^{AB} \nonumber
\eea
where $c_{\pm}>0$ and $P_{\pm}^{AB}\geq0$ are projections onto positive and negative eigenvalues, respectively. In a prepare-and-measure (P\&M) scenario, an experimenter realizes a measurement with $P_{\pm}^{AB}$ and then performs measurements to find $\tr[P_{\pm}^{AB} \rho^{AB}]$ to construct an expectation value $\tr[W^{AB}\rho^{AB}]$, see Equation \ref{eq:1}.

In cloud quantum computing, it is aimed to find if an entangled state $\rho$ can be generated among qubits labeled by $A$ and $B$ in the service. To this end, a user asks the service to prepare quantum circuits that realize a state $\rho$ and a measurement $P_{\pm}$ in qubits $A$ and $B$. Qubit allocations signify the service's mapping between the circuits by a user and the qubits labeled in quantum hardware. If a service addresses qubits $A$ and $B$ in the estimation of $\tr[P_{+} \rho]$ but some other ones for $\tr[P_{-} \rho]$, say $C$ and $D$, the service fails to estimate the expectation value $\tr[W\rho]$, i.e., entanglement generation between two qubits $A$ and $B$ of interest is not certified. We stress that it is crucial to ensure that the same qubits are addressed in the estimation of both quantities $\tr[P_{\pm} \rho]$: in this way, entanglement generation is in any case certified for two qubits that may not be identified. 

For instance, the IBMQ service provides qubit allocations in two ways, manual or automatic: a user can choose the qubits in the service or ask the service to decide those qubits used. In the latter case of automatic allocations, the service also reports those qubits used during the circuit evaluation. However, the command of qubit allocations is not always available: it lacks in the IonQ service, where different qubits are addressed at every time a new circuit is applied. Thus, EWs cannot be realized in IonQ, although they can be implemented in IBMQ under the assumption that qubit allocations by a service are trusted.

It is also worth mentioning that it is an additional assumption to trust qubit allocations by a cloud computing service. In a quantum laboratory, e.g., entanglement detection in trapped ions \cite{Haffner:2005vw, Barreiro:2013wa}, they are experimenters which control qubit allocations. 

\subsection*{Summary of the contribution}

In this work, we introduce entanglement-witnessing circuits (EWCs) by transferring entanglement witnesses (EWs) in a P\&M scenario to a quantum circuit. We show that the capability of generating entanglement of quantum circuits can be certified by EWCs without quantum tomography. Remarkably, the framework relaxes the assumption of qubit allocations. In other words, the validity of a conclusion in the certification framework does not rely on how successful or unsuccessful qubit allocations are in a cloud computing service. The results show that the framework of EWCs achieves a higher level of certification; as the number of assumptions is fewer, the higher the level of certification is obtained. 

\begin{center}
\begin{tabular}{l*{6}{c}r}
     Scenario & Preparation & Certification  \\ \hline
    \hline
    P \& M & States  & Entanglement \\ 
    Computing & Circuits   & Entanglement generation \\ 
\end{tabular}
\end{center}

Our approach to constructing EWCs relies on the structural physical approximation (SPA) \cite{PhysRevLett.89.127902, Bae_2017} that facilitates a transformation of EWs to non-negative operators that can be prepared as EWCs. Once EWs are transferred to EWCs, we employ the brand new and efficient framework called EW 2.0, which boosts the capability of an EW twice higher \cite{Bae:2020aa}. Compared to Equation \ref{eq:1}, EW 2.0 offers two bounds, satisfied by all separable states, to an EW. Then, entangled states can be concluded by a violation of either of the bounds. Consequently, EWCs constructed from EW 2.0 are twice more efficient compared to the cases followed by standard EWs. All these constitute the architecture of certifying EGCs in a circuit-based quantum computing model. As an EW is to the detection of entangled states, an EWC is to the certification of entanglement generation by a quantum circuit. 

We then apply EWCs to certify entanglement generation in cloud-based quantum computing services in the IBMQ and IonQ services. As mentioned above, a command of qubit allocations is provided in the IonQ service, for which our framework successfully certifies entanglement generation. Note that it is the first time that entanglement generation is certified in the IonQ hardware. We stress that it is EWCs that make it possible to certify entanglement generation in cases when qubit allocations are not trusted. The quantification of entanglement generation in IBMQ and IonQ is presented from the data collected in the certification.

The article is organized as follows. In Section \ref{sec:3}, entanglement detection in a P\&M scenario is presented. Examples of two- and three-qubit entangled states are shown as consequences of universal gates. The brand new framework EW 2.0 is also summarized. In Section \ref{sec:4}, the P\&M scenario is transferred to a circuit-based quantum computing model. The architecture for certifying EGCs is shown by constructing EWCs via SPA and EW 2.0. In Section \ref{sec:ibmq}, the certification of two-qubit EGCs is demonstrated in IBMQ Casablanca. In Section \ref{sec:ionq}, two- and three-qubit EGCs are realized in IonQ and their certification is demonstrated. In Section \ref{sec:discussion}, the results with IBMQ and IonQ systems are compared and discussed. In Section \ref{sec:conclusion}, we summarize the results and conclude their applications with NISQ technologies.

\section{ Entangled states}
\label{sec:3}

In this section, the P\&M scenario of detecting entangled states is summarized. The recent result of EW 2.0 is also reviewed with illuminating examples. 

\begin{figure*}[t]
	\begin{center}
		\includegraphics[angle=0, width=0.95\textwidth]{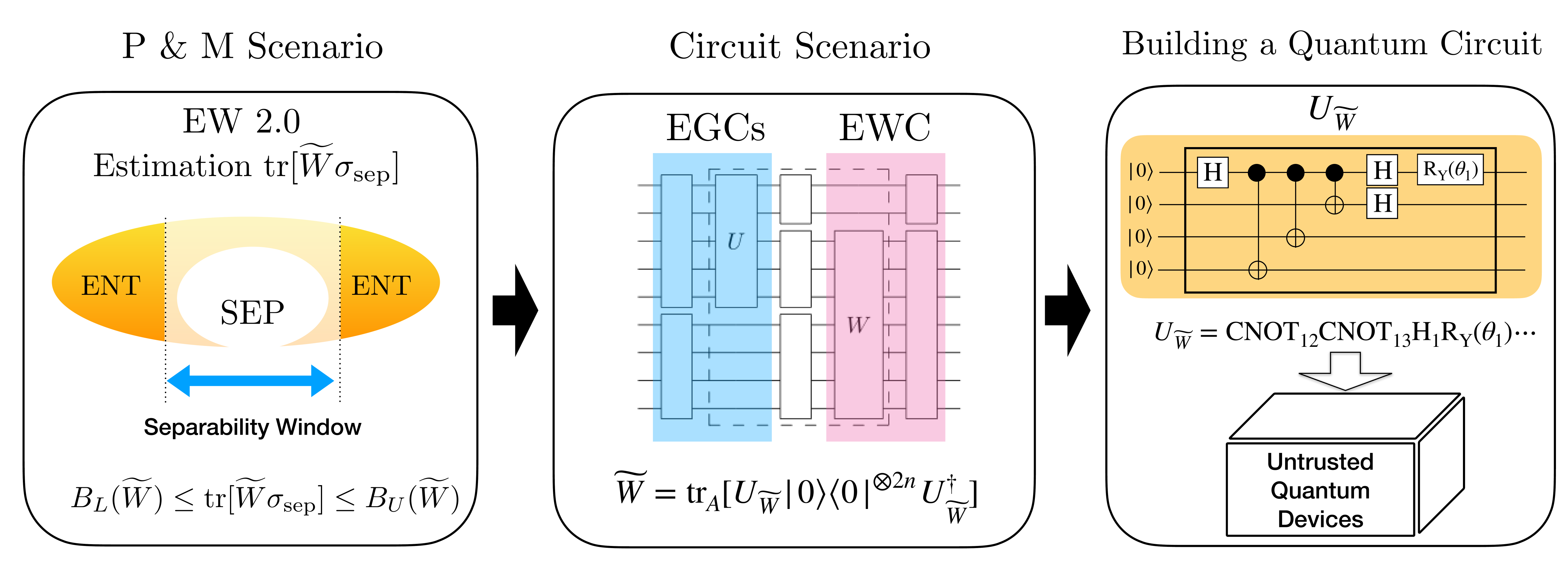}
		\caption{ The steps of designing an EWC that certifies an EGC are summarized. The design begins from EW 2.0 in a P\&M scenario. An EW 2.0 observable $\widetilde{W}$ twice more efficient than a standard EW is transferred to an EWC $U_{\widetilde{W}}$. The circuit construction of the EWC is optimized in terms of universal gates. The designed EWC is applied to certifying EGCs realized in untrusted quantum hardware.}  \label{figure}
	\end{center}	
\end{figure*}

\subsection{Entanglement and universal gates}

A quantum circuit can be generally decomposed into a set of quantum gates. Conversely, a collection of gates can efficiently approximate an arbitrary quantum circuit. The property is referred to as universality, and defines a set of universal gates \cite{PhysRevA.52.3457}. The efficiency in the approximation can be estimated via the Solovay-Kitaev theorem \cite{10.5555/863284}.  For instance, well known examples of universal gates are single-qubit and a controlled-NOT gates. Or, a Toffoli and single-qubit gates are also universal. Hence, it suffices to consider two- and three-qubit entangled states for the certification of EGCs. 

\begin{tcolorbox}
Entanglement-generating circuits (EGCs) : A circuit $U$ for $n$ qubits is an EGC if there is an $n$-qubit unentangled state $|\psi\rangle$ such that the resulting state $U|\psi\rangle$ is entangled. In general, an $n$-qubit EGC can be factorized into two- and three-qubit EGCs. 
\end{tcolorbox}

\subsection{ Entanglement witness 2.0 }


In general, an EW can be realized by local measurements and classical communication. This means that an EW $W$ admits a decomposition as follows,
\bea
W = \sum_{ij} c_{ij} M_i \otimes M_j \label{eq:ewlocc}
\eea
with a quantum measurement described by positive-operator-valued-measure (POVM) elements $\{ M_i\}$ and $\{M_j \}$, where $c_{ij}$ are real numbers. Equation \ref{eq:ewlocc} also tells how to relate measurement outcomes to entanglement detection. Once measurement outcomes are collected, one can find probability distributions, $p_{\mathrm{exp}}(ij) = \tr[\rho M_i \otimes M_j ]$ for a unknown state $\rho$. Then, the expectation value can be obtained by collecting the probabilities, 
\bea
 \tr[W\rho] =\sum_{ij} c_{ij} p_{\mathrm{exp}} ( ij). \label{eq:expew}
\eea
Negative expectation values verify entangled states. 

Let us illustrate how an EW can detect entangled states. The unit of entanglement is defined by an entangled bit (ebit) \cite{PhysRevLett.76.722}, 
\bea
\mathrm{ebit}:~ |\phi^+ \rangle =\frac{1}{\sqrt{2}} (| 00\rangle + |11\rangle) \label{eq:ebit}
\eea
which contains various usefulness, quantum teleportation \cite{PhysRevLett.70.1895}, quantum key distribution \cite{PhysRevLett.68.557}, measurement-based quantum computing \cite{PhysRevLett.86.5188}, etc. Let us consider a noisy ebit  
\bea
\rho = (1-p) |\phi^+\rangle \langle \phi^+| + \frac{p}{4} I\otimes I. \label{eq:iso}
\eea
where $I$ denotes the $2\times 2$ identity matrix. This state is entangled for $p\in [0,2/3)$ \cite{PhysRevLett.77.1413, HORODECKI19961}. An EW for noisy ebits can be constructed,
\bea
W = \frac{1}{2} ( I\otimes I + X\otimes X + Y\otimes Y -Z\otimes Z )  \label{eq:ewex}
\eea
where $X$, $Y$ and $Z$ are Pauli matrices
\bea
 X = \qgate{0}{1}{1}{0}, ~ Y = \qgate{0}{-i}{i}{0}, ~ Z = \qgate{1}{0}{0}{-1}. \label{eq:pauli}
\eea
Then, it hods that $\tr[W\rho]<0$ for all $p\in [0,2/3)$. This also shows that entanglement detection with EWs is robust to noise. 

Note, however, that EWs are not efficient: the EW in Equation \ref{eq:ewex} fail to detect other Bell states:
\bea
&& \tr[W|\phi^-\rangle \langle \phi^-|] >0, ~ \tr[W|\psi^+ \rangle \langle \phi^+ |] >0,\nonumber \\
&\mathrm{and}&~ \tr[W|\psi^-\rangle \langle \psi^-|] >0 \nonumber
\eea
where $|\phi^- \rangle = (I \otimes Z) |\phi^+ \rangle$, $|\psi^+ \rangle = (I \otimes X) |\phi^+ \rangle$ and $|\psi^- \rangle = (I \otimes Y) |\phi^+ \rangle$. To be precise, an optimal EW for a single Bell state cannot detect other Bell states. 

The framework of EW 2.0 improves the detection capability of a single EW \cite{Bae:2020aa}.  Let us begin with an EW for two-qubit states, denoted by $W^{(+)}$. Its positive structural physical approximation (SPA) has been introduced in \cite{PhysRevA.68.052101, PhysRevLett.89.127902} as a mapping from $W^{(+)}\ngeq 0$ to $\widetilde{W}\geq 0$
\bea
\widetilde{W} = (1-p) W^{(+)} + p\frac{1}{4} I\otimes I  \label{eq:pSPA}
\eea
with a minimal $p\in(0,1)$. We call the construction as a positive SPA in that $(1-p)>0$. A negative SPA can be defined with $(1-p)<0$ in the other way around such that the resulting operator is non-negative. Let us then deduce an EW $W^{(-)}$ from the SPAed one $\widetilde{W}$ above via negative SPA as follows
\bea
\widetilde{W} = (1-q) W^{(-)} +q  \frac{1}{4} I\otimes I  \label{eq:nSPA}
\eea
with a maximal $q>1$. Then, we call $W^{(-)}$ is mirrored to $W^{(+)}$, vice versa. Note that two EWs are obtained from a single SPAed non-negative operator $\widetilde{W}$. That is, given an EW, EW 2.0 tells us how to crop an additional EW.

In Ref. \cite{Bae:2020aa}, it has been shown that a SPAed operator $\widetilde{W}$ can contain the detection capabilities of both of the mirrored EWs. This can be seen by lower and upper bounds satisfied by all separable states, called EW 2.0.

\begin{tcolorbox}
The framework of EW 2.0:
\bea
\forall \sigma_{\mathrm{sep}},~~ B_L ({\widetilde{W}}) \leq \tr[\widetilde{W}\sigma_{\mathrm{sep}}] \leq B_U( {\widetilde{W}} )\label{eq:bound}
\eea
where 
\bea
B_L ( {\widetilde{W}} ) &=& \min_{\sigma_{\mathrm{sep}}} \tr[\widetilde{W}\sigma_{\mathrm{sep}}] ~~\mathrm{and}~~ \nonumber\\
B_U ( {\widetilde{W}} ) &=&\max_{\sigma_{\mathrm{sep}}} \tr[\widetilde{W}\sigma_{\mathrm{sep}}].  \nonumber
\eea
The range $[B_L ( {\widetilde{W}} ), B_U ( {\widetilde{W}} )]$ is called the separability window for $\widetilde{W}$.
\end{tcolorbox}

The set of entangled states violating the lower bound is the same to that of those states detected by $W^{(+)}$, i.e., the lower bound is equivalent to the condition that $\tr[W^{(+)} \sigma_{\mathrm{sep}}]\geq 0$ for all separable states $\sigma_{\mathrm{sep}}$. Similarly, the upper bound is equivalent to the condition with its mirrored one $W^{(-)}$ that $\tr[W^{(-)} \sigma_{\mathrm{sep}}] \geq 0$ for all separable states $\sigma_{\mathrm{sep}}$. We stress that as it is shown in Equation \ref{eq:bound}, two EWs are implemented in the estimation of a single observable $\widetilde{W}$.
 
Two remarks are in order as follows. Firstly, it is important to find probabilities $p_{\mathrm{exp}} ( ij)$ for all $i,j$ in Equation \ref{eq:expew} from experimental data. This means that a designed quantum circuit needs to be repeated sufficiently many times for different POVM elements. Secondly, it is crucial to assume that a measurement with POVM elements  $\{ M_i\}$ and $\{M_j \}$ in Equation \ref{eq:expew} is performed on the qubits for which it is questioned whether their state is entangled, or not; otherwise, a measurement setup that prepares an EW would fail in the certification. In cloud-based quantum computing, this is translated to the assumption that a qubit allocation is trusted. If a qubit allocation is not provided in the service such as IonQ, EWs cannot be directly applied to detecting entangled states. 

\subsection{Two-qubit entanglement}

In this subsection, we show how to construct EW 2.0 for two-qubit entangled states. For two-qubit states, a controlled-NOT gate that generates Bell states constitutes a universal set of quantum gates. We present how to construct EW 2.0 for detecting multiple Bell states, whereas a standard EW can detect a single Bell state only. 

Let us consider with two Bell states $|\phi^+\rangle$ and $|\psi^-\rangle$. It has been known that an EW $W^{(+)}$ in the following can be applied to detect the state $|\phi^+\rangle$ \cite{PhysRevLett.94.060501, PhysRevA.72.022340, PhysRevA.101.012317},
\begin{align}
    W^{(+)} &= \frac{1}{4} I\otimes I - \frac{1}{2} \left(  |\phi^+\rangle \langle \phi^+| -  |\psi^- \rangle \langle \psi^-|  \right) \\
        &= \frac{1}{4} \left( \begin{array}{cc|cc}
            0 & 0 & 0 & -1  \\
            0 & 2 & -1 & 0  \\
             \hline
            0 & -1 & 2 & 0  \\
            -1 & 0 & 0 & 0
        \end{array} \right), \label{eq:ew2p}
\end{align}
One can obtain its mirrored EW $W^{(-)}$ via EW 2.0,
\begin{align}
    W^{(-)} &= \frac{1}{4} I + \frac{1}{2} \left(   |\phi^+\rangle \langle \phi^+| -  |\psi^- \rangle \langle \psi^-| \right) \\
        &= \frac{1}{4}  \left( \begin{array}{cc|cc}
            2 & 0 & 0 & 1  \\
            0 & 0 & 1 & 0  \\
             \hline
            0 & 1 & 0 & 0  \\
            1 & 0 & 0 & 2
        \end{array} \right), \label{eq:ew2m}
\end{align}
which detects $| \psi^-\rangle$, i.e., $\tr (W^{(-)}   |\psi^- \rangle \langle \psi^-| ) = -1/4$. Note that the EWs are robust to noise, as they also work when the Bell states are admixed with a depolarization. It follows that the SPAed operator is constructed: 
\bea
    \widetilde{W} & = &  (1- p) W^{(+)} + p  \frac{1}{4}I\otimes I, \nonumber \\
    & = & (1-q) W^{(-)} + q  \frac{1}{4}I\otimes I \label{eq:spaed2}
\eea
with $p=1/2$ and $q=3/2$. It is straightforward to construct the inequalities satisfied by all separable states $\sigma_{\mathrm{sep}}$,
\bea
\frac{1}{8} \leq \tr[\widetilde{W} \sigma_{\mathrm{sep}}] \leq \frac{3}{8} \label{eq:ew2}
\eea
where the lower and upper bounds are equivalent to detection by $W^{(+)}$ and $W^{(-)}$, respectively. Thus, the framework of EW 2.0 is shown for detecting two Bell states.

\subsection{Three-qubit entanglement}

As a three-qubit gate, a Toffoli gate, generating three-qubit entangled states constitutes a universal set of quantum gates, let us show how to construct EW 2.0 for detecting three-qubit entangled states. We recall that three-qubit entangled states are classified into two classes, Greenberger-Horne-Zeilinger (GHZ) \cite{Greenberger:1989wc} and W types, which cannot be connected by stochastic local operations and classical communication. It turns out that almost all three-qubit entangled pure states are of the GHZ type \cite{PhysRevA.62.062314}. Hence, let us consider an EW in the following for detecting a GHZ state \cite{PhysRevLett.94.060501, PhysRevA.72.022340, PhysRevA.101.012317},
\bea
    W^{(+)} &=& \frac{1}{8} \left( I_1 I_2 I_3 - \frac{1}{2}(X_1 X_2 X_3 + Z_1 Z_2 + Z_2 Z_3) \right)~~ \label{eq:ew3p} 
\eea
where we have used the convention that labels the qubit in the subscript, e.g., $I_1 I_2 I_3 = I\otimes I \otimes I$ and $Z_2 Z_3 = I\otimes Z \otimes Z$. It is useful to rewrite the EW in terms of projectors to find those states detected, 
\bea
&&   W^{(+)} = \nonumber \\
    && \frac{1}{8} \left( -\frac{1}{2} |\phi^+_{000}\rangle \langle \phi^+_{000} | + \frac{1}{2}  |\phi^-_{000}\rangle \langle \phi^-_{000}| + \frac{1}{2} |\phi^+_{001}\rangle \langle \phi^+_{001}| \right. \nonumber \\
    & & \left. + \frac{3}{2} |\phi^-_{001}\rangle \langle \phi^-_{001}| + \frac{3}{2}  |\phi^+_{010}\rangle \langle \phi^+_{010}| + \frac{5}{2}  |\phi^-_{010}\rangle \langle \phi^-_{010}| \right. \nonumber \\
    && \left. + \frac{1}{2} |\phi^+_{011}\rangle \langle \phi^+_{011}| + \frac{3}{2} |\phi^-_{011}\rangle \langle \phi^-_{011}| \right), \label{eq:ew3p2}
\eea
where $\ket{\phi^{\pm}_{\x}} =  (\ket{\x} \pm \ket{\bar{\x}})/\sqrt{2}$ for a sequence $\mathrm{x} = \x_1 \x_2 \x_3$ and $\bar{\x} = \bar{\x}_1 \bar{\x}_2 \bar{\x}_3$ with $\bar{\x}_1 =  1-\x_1$. For instance, a GHZ state $\ket{\phi^+_{000}} =  (\ket{000} + \ket{111}) /\sqrt{2}$ is detected: one can find that $\tr [ W^{(+)} | \phi^+_{000}\rangle \langle \phi^+_{000}| ] =- 1/16$.

Its mirrored EW $W^{(-)}$ is then obtained via EW 2.0,
\bea
    W^{(-)} &=& \frac{1}{8} \left( I_1 I_2 I_3 + \frac{1}{2}(X_1 X_2 X_3 + Z_1 Z_2 + Z_2 Z_3) \right). ~~\label{eq:ew3m}
    \eea
It is also useful to rewrite it in terms of projectors to find those entangled states detected by the EW,
\bea
&&    W^{(-)} = \nonumber\\
    && \frac{1}{8} \left( \frac{5}{2} |\phi^+_{000}\rangle \langle \phi^{+}_{000}| + \frac{3}{2} |\phi^-_{000}\rangle\langle\phi^-_{000}| + \frac{3}{2} |\phi^+_{001}\rangle \langle \phi^+_{001}| \right. \nonumber \\
    & & \left. + \frac{1}{2} |\phi^-_{001}\rangle \langle \phi^-_{001}| + \frac{1}{2} |\phi^+_{010}\rangle \langle \phi^+_{010}| - \frac{1}{2} |\phi^-_{010}\rangle \langle \phi^-_{010}| \right. \nonumber \\
    && \left. + \frac{3}{2} |\phi^+_{011}\rangle \langle \phi^+_{011}| + \frac{1}{2} |\phi^-_{011}\rangle \langle \phi^-_{011}| \right). \label{eq:ew3m2}
\eea
It is clear that EW $W^{(-)}$ can detect another type of a GHZ state $\ket{\phi_{010}}$, i.e., $\tr [ W^{(-)} |\phi^-_{010} \rangle \langle \phi^-_{010}| ]= -1/16$.

The framework of EW 2.0 is completed by a single non-negative observable that contains both EWs $W^{(\pm)}$,
\bea
    \widetilde{W} &=& \frac{1}{8} \left( I_1 I_2 I_3- \frac{1}{2}(X_1 X_2 X_3 + Z_1 Z_2 + Z_2 Z_3) \right). \label{eq:spa3ew}
\eea
To clarify the he relations with the EWs, one can find the identities in the following
\bea
    \widetilde{W} & = &  (1-p) W^{(+)} + p  \frac{1}{8} I_1 I_2 I_3  \nonumber \\
    & = & (1-q) W^{(-)} + q  \frac{1}{8}  I_1 I_2 I_3  \nonumber
\eea
with $p=1/3$ and $q=5/3$. After all, the framework of EW 2.0 is shown with inequalities satisfied by all separable states,
\bea
\frac{1}{24} \leq \tr[\widetilde{W} \sigma_{\mathrm{sep}}] \leq \frac{5}{24} \label{eq:ew3}
\eea
It is shown how EW 2.0 can be constructed to detect GHZ-type three-qubit entangled states.

\section{ Entanglement Generating Circuits}
\label{sec:4}

 Quantum computing is structured by initialization, a quantum dynamics, e.g., a quantum circuit, and a measurement. The initialization prepares $n$ qubits in a fixed state $|0\rangle^{\otimes n}$. Then, a designed quantum circuit is performed, which corresponds to a quantum algorithm. A measurement is performed in the computational basis $|0\rangle$ and $|1\rangle$. The outcome statistics is used to conclude a computational task.  

The goal is to certify entanglement generation in a quantum circuit. Let $U_G$ denote a quantum circuit realized in quantum computing. Entanglement generation may be certified by applying EW 2.0 to a resulting state $U_G |0\rangle^{\otimes}$. We then have to set the framework of transferring EWs to a circuit architecture for the certification of entanglement generation by $U_G$. As mentioned earlier, it suffices to consider circuits $U_G$ of two and three qubits. Apart from single-qubit gates, a set of universal quantum gates contains either two- or three-qubit gates.  

In this section, we show how to construct entanglement-witnessing circuits (EWCs) via the framework of EW 2.0. It is also shown that, by doing so, qubit allocations can be relaxed.

\subsection{ Building a quantum circuit }


An arbitrary quantum circuit can be efficiently approximated by a set of universal gates, which means that single- and two-qubit gates are universal \cite{10.5555/863284}. Note also that Toffoli and Hadamard gates are universal, where a Toffoli gate is a three-qubit entangling gate. These hugely simply the certification scenario into two- and three-qubit EGCs.

In the case of single-qubit gates, there is a general decomposition in terms of three rotations. A single qubit gate $U$ can be implemented as $U = e^{i\alpha} AXBXC$ where $A$, $B$, and $C$ satisfies $ABC=I$ and $\alpha$ is a phase \cite{nielsen_chuang_2010}. For instance, one can choose 
\bea
A & = & R_z(\beta) R_y(\gamma/2), \nonumber \\
B & = & R_y(-\gamma/2)  R_z(-(\delta+\beta)/2), \nonumber \\
C & = & R_z((\delta-\beta)/2),\nonumber
\eea
where $R_{\hat{n}} (\theta)$ denotes a rotation counterclockwise by $\theta$ about the axis $\hat{n}$ in the Bloch sphere. A two-qubit gate can be written as a controlled-$U$ gate for a single-qubit gate $U$,
\bea
|0\rangle\langle 0| \otimes I +  |1\rangle\langle 1| \otimes U. \nonumber
\eea
Throughout, we use the gate decomposition introduced above.

\begin{figure}[h]
	\begin{center}
		\includegraphics[angle=0, width=0.48\textwidth]{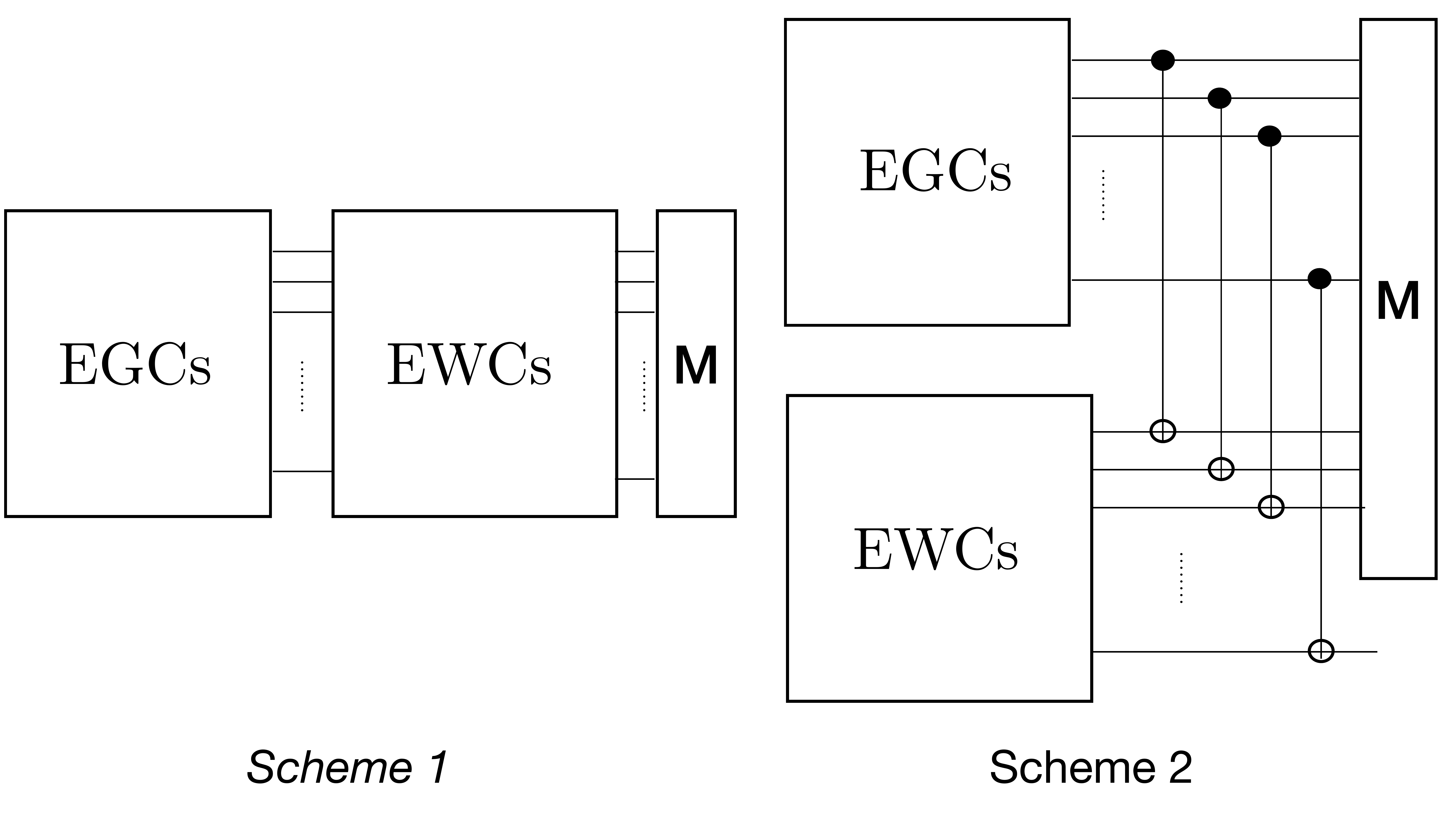}
		\caption{ Scheme 1 and Scheme 2 for certifying EGCs are presented. In Scheme 1, an EWC is applied right after an EGC is applied, and certifies an EGC. A two-qubit EWC is also shown in Figure \ref{2qubitibmq2}. A drawback of the scheme is that the overall circuit depth increases by both of the circuits. This can be prescribed in Scheme 2 where two circuits are prepared in parallel. The overall circuit depth is then as much as that of an EGC addition to the number of qubits. 
		 } \label{pic2}
	\end{center}	
\end{figure}

\subsection{ The architecture for certifying EGCs }

The framework of EW 2.0 makes it possible to compress that multiple EWs into estimation of a single observable $\widetilde{W}$. The observable is nothing but a non-negative and unit-trace operator; it is important that its purification can be admitted. Namely, there exists a state $|\widetilde{W}\rangle_{SA}$ over a system $S$ and an ancilla $A$ such that:
\bea
\widetilde{W} = \tr_A | \widetilde{W}\rangle_{SA} \langle \widetilde{W} |  \label{eq:puri}
\eea
Note that a standard EW $W\ngeq 0$ does not admit a purification. 

Then, the central idea of certifying $n$-qubit EGCs is to estimate the expectation value of an SPAed operator $\widetilde{W}$ via its purification. Let us reiterate the formalism of EW 2.0 in Equation \ref{eq:bound} in terms of quantum circuits. For an unknown $n$-qubit pure state $\rho = U_G |0\rangle\langle 0|^{\otimes n}U_{G}^{\dagger} $, the estimation is identified by
\bea
\tr[\widetilde{W} \rho] 
& = & \tr[  | \widetilde{W}\rangle_{SA} \langle \widetilde{W} |  ~\rho_S\otimes I_A ]  \nonumber\\
&=& \tr[  | \widetilde{W}\rangle_{SA} \langle \widetilde{W} |  ~ U_G |0\rangle_S \langle0|^{\otimes n} \otimes I_ A U_{G}^{\dagger} ]   \label{eq:bound2}
\eea
where $U_G$ denotes a quantum circuit that generates a state $\rho$: thus, $U_G$ is also unknown. Note that from the framework of EW 2.0, on can conclude that a preparation circuit $U_G$ is an EGC if the expectation in Equation \ref{eq:bound2} is not within the lower and upper bounds in Equation \ref{eq:bound}.

It is left to construct EWCs that certifies EGCs. In what follows, two schemes of constructing EWCs are presented.

\begin{figure*}[t]
	\begin{center}
		\includegraphics[angle=0, width=0.9\textwidth]{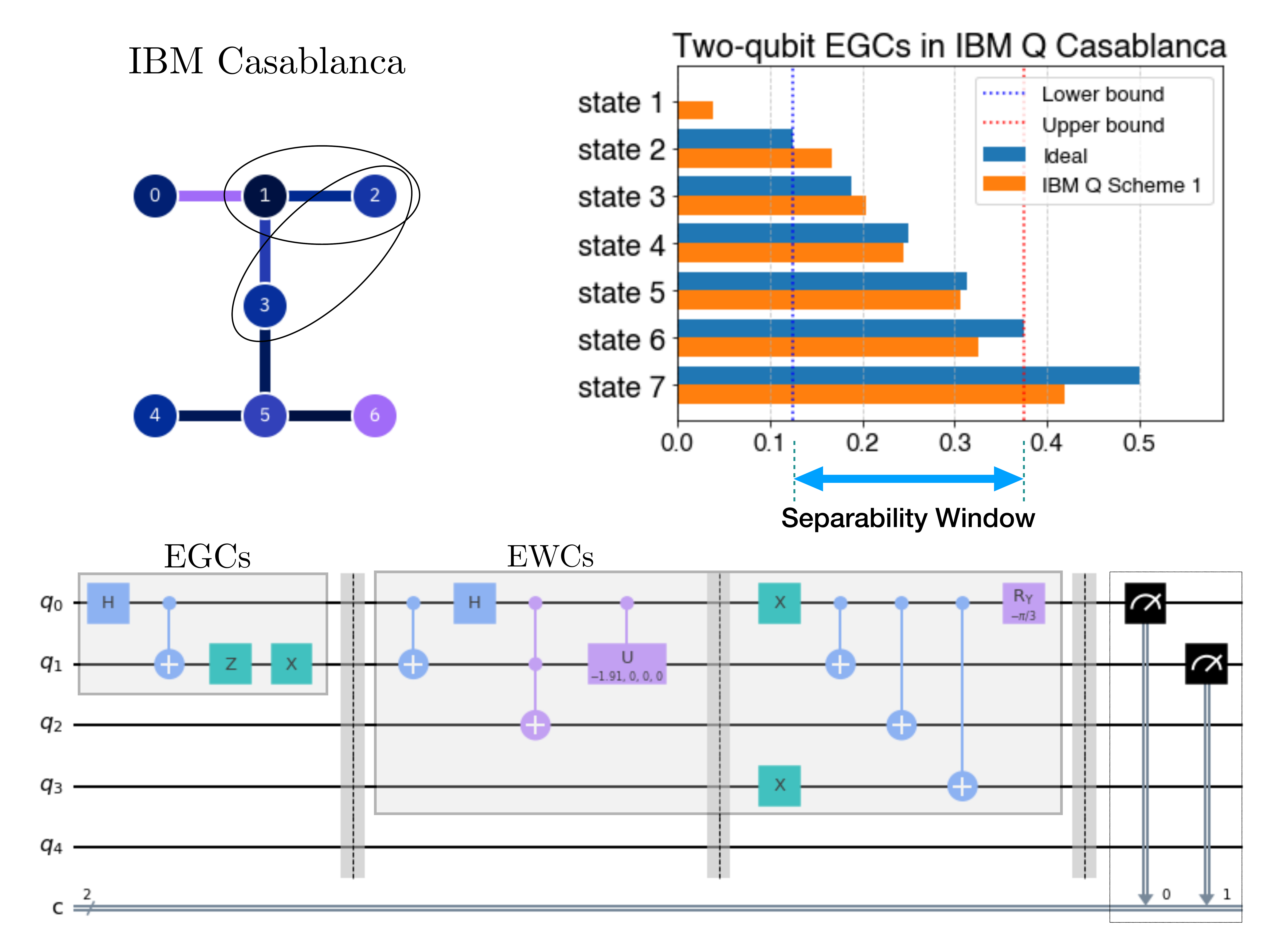}
		\caption{ Entanglement generation is certified in two qubits labeled $(1,2)$ and $(2,3)$ in IBMQ Casablanca. For a pair of the qubits, two-qubit circuits that generate entangled and separable states in Equation \ref{eq:2state} are performed. An EWC is constructed from the framework of EW 2.0 is then applied to certify entanglement generation. The cloud-based quantum computing service returns the outcome statistics. EGCs are certified if either the lower or the upper bound is violated. The separability window is given by $[0.125, 0.375]$: estimated values off the range certify entangled states. EGCs generating state 1 and state 7 in Equation \ref{eq:2state} are certified. The command of automatic qubit allocations is applied in the certification. }  \label{2qubitibmq2}
	\end{center}	
\end{figure*}

\subsubsection{ Scheme 1 }

One can find a unitary transformation $U_{\widetilde{W}}$ that prepares a purification $ |\widetilde{W}\rangle_{SA}$, see Equation \ref{eq:puri}
\bea
|\widetilde{W}\rangle_{SA} = U_{\widetilde{W}} |0\rangle^{\otimes 2n} \label{eq:uw}
\eea
where $2n$ qubits are sufficient for the purification. As the circuit $U_{\widetilde{W}}$ is used to detect entangled states, i.e., to find if a circuit $U_G$ is an EGC, it is an EWC. With an EWC $U_{\widetilde{W}}$, it holds that
\bea
&& p(0\cdots 0) =  \tr[\widetilde{W} \rho]  \nonumber \\
 &=&\tr_{SA} ~[ |0\rangle\langle 0|^{\otimes 2n} ~U_{\widetilde{W}}^{\dagger} U_G   |0\rangle\langle 0|^{\otimes n} \otimes I_A ~ U_{G}^{\dagger} U_{\widetilde{W}}] ~~~\label{eq:000}
\eea
where $p(0\cdots0)$ denotes the probability of obtaining the $2n$-bit sequence $0\cdots 0$ in a measurement. 

Let us explain Equation \ref{eq:000} as follows. First, $n$ qubits are initialized in the state $|0\rangle^{\otimes n}$ and a circuit $U_G$ that aims to generate an $n$-qubit entangled state is applied. Note that Equations \ref{eq:bound2} and \ref{eq:000} are identical, where we have used the relation that $\tr[AB] = \tr[BA]$ for two operators $A$ and $B$. Then, right after $U_G$, the certification circuit $U_{\widetilde{W}}^{\dagger}$ constructed from an SPAed operator in Equation \ref{eq:uw} is applied to find if $U_G$ is an EGC. Measurements on $2n$ qubits are performed in the computational basis. The probability of obtaining $2n$-bit sequence $0\cdots 0$ in the measurement corresponds to the estimation in the framework of EW 2.0. If the probability is not within the upper and lower bounds in Equation \ref{eq:bound}, an EWC $U_{\widetilde{W}}$ certifies that $U_G$ is an EGC. 

\begin{tcolorbox}
The construction of EWCs in Scheme 1: Consider $n$-qubit state $\rho = U_G |0\rangle \langle 0|^{\otimes n} U_{G}^{\dagger}$ and EW 2.0 in terms of a single obervable $\widetilde{W} = \tr_A | \widetilde{W}\rangle_{SA} \langle \widetilde{W} |$ where $|\widetilde{W}\rangle_{SA} = U_{\widetilde{W}} |0\rangle^{\otimes 2n}$. Entanglement of the state $\rho$ is detected by $\widetilde{W}$. Then, the circuit $U_{\widetilde{W}}$ can be used as an EWC to certify entanglement generation by $U_G$.
\end{tcolorbox}

\subsubsection{Scheme 2}

The probability in Equation \ref{eq:000} may be realized in a shorter-depth scheme. The strategy is to translate the estimation scheme in the previous section in a measurement-based way. Let us first show how an overlap of two states can be estimated. Let  $\sigma_1$ and $\sigma_2$ denote single qubit states where it is supposed that the state $\sigma_2$ is known. The Hilbert-Schmidt product can be computed as follows \cite{pillis, Jamiokowski:1972tf, Choi:1975tv, PhysRevLett.86.544},
\bea
\tr[\sigma_1 \sigma_2 ] =  \tr[\sigma_1 \otimes \sigma_{2}^T ~P_{12} ] \nonumber
\eea
where $P_{12}  = 2 |\phi_{}^{+}\rangle_{12} \langle \phi_{}^{+}|$. Since $\sigma_2$ is a quantum state, its transposed one $\sigma_{2}^{T}$ also corresponds to a state. Since the state is known, one can also prepare its transposed state. Then, let $V_{12}$ denote an entangling circuit that prepares an ebit, i.e., $P_{12} = 2 V_{12}|00\rangle \langle 00| V_{12}^{\dagger}$. One can rewrite the identity above,
\bea
\tr[\sigma_1 \sigma_2 ] = 2 p(00) \label{eq:r1}
\eea
where
\bea
p(00) = \tr[ V_{12}^{\dagger} \sigma_1 \otimes \sigma_{2}^T V_{12} ~ |00\rangle \langle 00| ] \label{eq:r2}
\eea
is the probability of obtaining the measurement outcome $00$ after two qubit states $ \sigma_1 \otimes \sigma_{2}^T$ interact with each other via the unitary $V_{12}^{\dagger}$. Note that a controlled-NOT (CNOT) gate is contained in the interaction $V_{12}$.

Applying the identity above, one can rewrite Equation \ref{eq:bound2} for an $n$-qubit state $\rho_S$ with a set $S$ collecting $n$ qubits labeled by $1\cdots n$, as follows
\bea
&& \tr [ \widetilde{W}_{1\cdots n} \rho _{1\cdots n}]  \nonumber \\
&=& \tr  [\rho _{1\cdots n}  \otimes \widetilde{W}_{1^{'} \cdots n^{'}}^T  ~P_{11^{'}}\otimes \cdots \otimes P_{nn^{'}}  ] \nonumber \\
&=& \tr  [U_G |0\rangle_S \langle 0|^{\otimes n}U_{G}^{\dagger}  \otimes ( U_{\widetilde{W}^T} |0\rangle_{S^{'}A} \langle 0|^{\otimes 2n} U_{\widetilde{W}^T}^{\dagger} )~ P_{SS^{'}}], ~~~~~~\label{eq:s2} 
\eea
where we have used that $P_{SS^{'}}$ is a projection over $2n$ qubits of system
\bea
P_{SS^{'}}= P_{11^{'}}\otimes \cdots \otimes P_{nn^{'}} \nonumber
\eea
and the certification circuit $U_{\widetilde{W}^T}$ prepares a purification of the transposed SPAed operator $\widetilde{W}_{}^T$, i.e., 
\bea
| \widetilde{W}_{}^T\rangle_{S^{'}A} = U_{\widetilde{W}^T} |0\rangle^{\otimes 2n}. \label{eq:cc2}
\eea
The $n$-ancila qubits denoted by $A$ are needed to prepare the $n$-qubit operator $\widetilde{W}_{}^T$. Note that each project $P_{k k^{'}}$ for $k=1,\cdots,n$ is constructed as, $P_{k k^{'}} =2 V_{kk^{'}} |00\rangle \langle00| V_{k k^{'}}^{\dagger}$. 

Consequently, Equation \ref{eq:s2} can be obtained by the probability in the following,
\bea
\tr [ \widetilde{W}_{1\cdots n} \rho _{1\cdots n}] = 2^n p(0\cdots 0) \label{eq:s2f}
\eea
where $p(0\cdots 0)$ denotes the probability of obtaining the $2n$-bit sequence $0\cdots 0$ in the registers $SS^{'}$,
\bea
&& p(0\cdots 0) =  \nonumber \\
&& \tr  [(V_{SS^{'}}^{\dagger} U_G    \otimes  U_{\widetilde{W}^T} |0\rangle \langle 0|^{\otimes 3n} U_{\widetilde{W}^T}^{\dagger} \otimes U_{G}^{\dagger} V_{SS^{'}} )~ |0\rangle_{SS^{'}}\langle 0|^{\otimes 2n} ]. \nonumber
\eea
Let us explain the expression in the above. Each register $S$, $S^{'}$ and $A$ contains $n$ qubits, where a preparation circuit $U_G$ and the certification circuit $U_{\widetilde{W}}^T$ are prepared on $S$ and $S^{'}A$, respectively. A known circuit $U_{\widetilde{W}}^T$ obtained from SPA to an EW $W$, see Equation \ref{eq:uw}, aims to find if a circuit $U_G$ generating an $n$-qubit state is an EGC. After both circuits are performed, the projection denoted by $V_{SS^{'}}$ is performed on registers $SS^{'}$, on which measurements are performed in the computational basis. Then, the probability of obtaining $2n$-bit $0\cdots 0$ concludes the expectation value in Equation \ref{eq:s2f}.

\begin{tcolorbox}{}
The advantage of Scheme 2: An EWC and an EGC are prepared independently. Then, controlled-NOT gates on $n$ qubits are performed to realize the certification of an EGC. The circuit depth increases linearly with the number of qubits. 
\end{tcolorbox}

\subsubsection{Comparison:}
Two schemes are equivalent in that both are realized by EW 2.0. On the side of resources, $Scheme~ 1$ exploits $2n$ qubits to implement an EWC and then performs measurements on $n$ qubits. Note that $Scheme~ 2$ requires more resources of $3n$-qubit for EWCs and EGCs. From the view of NISQ technologies, $Scheme ~2$ contains the projection step that requires two-qubit gates and measurements on more qubits, which may cause more errors. An advantage of $Scheme~2$ may appear when the circuit depth of an EWC is large: in this case, the circuit depth of $Scheme~1$ takes both an EWC and an EGC into account.

\begin{tcolorbox}
\textbf{The certification of EGCs}

Let $p_{ U_{\widetilde{W} } }( 0^{\times 2n}) [U]$ denote the probability of obtaining measurement outcomes $2n$ $0$'s after an application of an EWC $U_{\widetilde{W} }$ for a circuit $U$:
\bea
p_{U_{\widetilde{W}}} ( 0^{\times 2n})[U] = \tr[ |0\rangle \langle 0|^{\otimes 2n } U_{\widetilde{W}}^{\dagger} U_G   |0\rangle \langle 0|^{\otimes n } U_{G}^{\dagger} U_{\widetilde{W}}]. \nonumber
\eea
Then, the probability has has upper and lower bounds $B_L (\widetilde{W} )$ and $B_U (\widetilde{W} )$, respectively, for all quantum circuits $U_G$ that do not generate entanglement,
\bea
B_L (\widetilde{W} ) \leq p_{U_{\widetilde{W} }}( 0^{\times 2n})  [U_G]\leq B_U (\widetilde{W} ). \nonumber
\eea
When the inequalities are violated for an unknown circuit $U_G$, the circuit is certified to be an EGC.

We remark that the assumption of qubit allocations is not needed. Once the probability is obtained, one can conclude that an unknown circuit is an EGC.
\end{tcolorbox}

\subsubsection{Qubit allocations:}
We remark that the expectation value in Equation \ref{eq:bound2} is estimated by applying a unitary transformation, i.e., at one go in a quantum circuit. Since the initial state in a circuit is given by $|0\rangle^{\otimes n}$ it is not required to address which qubits are to be used. This immediately gets rid of the necessity of a qubit allocation to certify EGCs: no matter how a designed circuit is mapped to actual qubits in a cloud-based quantum computing service, violations of the bounds in EW 2.0 certify the presence of entanglement. Therefore, the assumption that a qubit allocation is trusted can be cleared out in the certification. Note that the assumption cannot be relaxed in standard EWs which do not have a purification. It is possible to clear the assumption by exploiting the framework EW 2.0.

\begin{figure*}[t]
	\begin{center}
		\includegraphics[angle=0, width=1\textwidth]{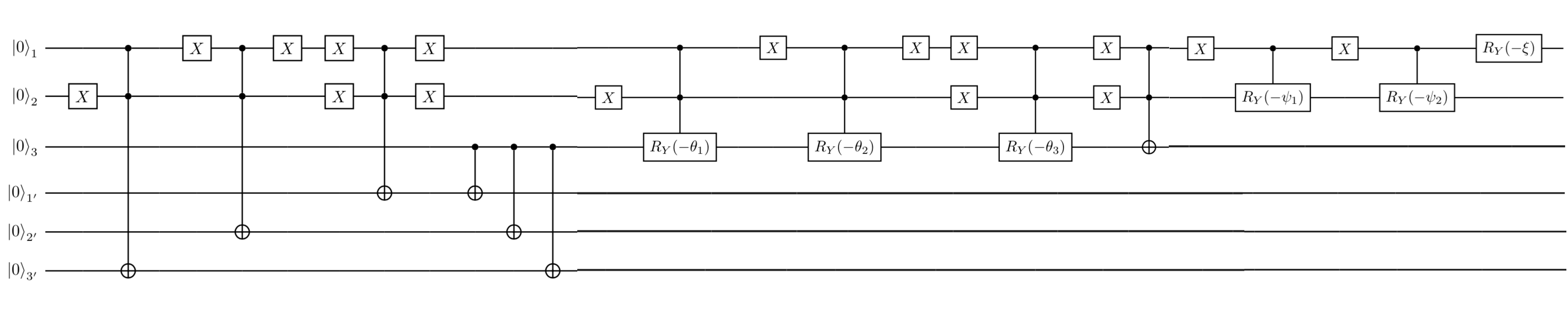}
		\caption{ An EWC is constructed from a unitary transformation generating the purification in $|\widetilde{W}\rangle$ in Equation \ref{eq:uw3}, where single-qubit gates are listed out in Equation \ref{eq:3gate}. The EWC is realized in IonQ and certifies entanglement generation. 	}  \label{3circuit}
	\end{center}	
\end{figure*}

\section{Demonstration in IBMQ }
\label{sec:ibmq}

The IBMQ system is based on superconducting qubits that have fixed locations and interact with each other remotely by concatenating SWAP gates. Note that IBMQ devices report errors on single-qubit gates around $0.1\%$, two-qubit gates and measurement readout errors around a few percentages. While two remote qubits interact with each other, the steps of SWAP may also cause errors. Then, a manual qubit allocation is provided: a user can address which qubits are used in a given geometry of qubits. If a qubit allocation is automatically applied, it is reported which qubits are used during the process at the end of runs of a quantum circuit. 

We apply the circuit architecture to the certification of entanglement generation in IBMQ. Two-qubit EWCs are realized and certify EGCs that generate Bell states. In Figure \ref{2qubitibmq2}. the arrangement of seven qubits and the circuits are shown. An EWC is constructed from the EW 2.0 operator,
\bea
\widetilde{W} _{12}= \frac{1}{4} |\phi^{-} \rangle_{12} \langle \phi^{-}|  +  \frac{1}{4} |\psi^{ + } \rangle_{12} \langle \psi^{ + }|    +   \frac{1}{2} |\psi^{ - } \rangle_{12} \langle \psi^{ - }|  \nonumber
\eea
that has the bounds $1/8=0.125$ and $3/8=0.375$, see also Equation \ref{eq:ew2}. Its EWC $U_{\widetilde{W}}$ can be obtained as follows,
\bea
&& | \widetilde{W} \rangle_{121^{'}2^{'}} =  U_{\widetilde{W}} |0\rangle_{121^{'}2^{'}}^{\otimes4} \nonumber \\
& = & \sqrt{ \frac{1}{4}}  |\phi^{-} \rangle | 01 \rangle +  \sqrt{ \frac{1}{4} } |\psi^{ + } \rangle |10\rangle    +   \sqrt{ \frac{1}{2}} |\psi^{ - } \rangle |11\rangle. ~~~~~~~ \label{eq:4ewpure}
\eea
The proof-of-principle demonstrations of EWCs are performed in IBM Casablanca on Jan. 22 2021 for quantum circuits $U_G$ generating the following states:
\bea
\mathrm{state ~1}: & ~~ &  |\phi^+\rangle =(|00\rangle + |11\rangle)/\sqrt{2} \nonumber \\
\mathrm{state ~2}:  &~~ & |00\rangle \langle 00|  \nonumber \\
\mathrm{state ~3}:  & ~~ &  (3/4)|00\rangle \langle 00| + (1/4)|01\rangle \langle 01| \nonumber \\
\mathrm{state ~4}: &~~  & (1/2)|00\rangle \langle 00| + (1/2)|01\rangle \langle 01| \nonumber \\
\mathrm{state ~5}:  & ~~ &  (1/4)|00\rangle \langle 00| + (3/4)|01\rangle \langle 01| \nonumber \\
\mathrm{state ~6}:  &~~  &  |01\rangle \langle 01| \nonumber \\
\mathrm{state ~7}:  & ~~ &  |\psi^-\rangle  =(|01 \rangle - |10\rangle)/\sqrt{2},\label{eq:2state}
\eea
where states 1 and 7 are entangled and the others are separable.

Note that a circuit to prepare mixed states, state $3$, $4$, and $5$, considers their purification by using three qubits with one ancilla qubit. For instance, state $5$ is obtained by preparing the following state,
\bea
|\phi_5\rangle = \sqrt{\frac{1}{4}} |00\rangle |0\rangle +  \sqrt{\frac{3}{4}} |01\rangle |1\rangle \nonumber
\eea
and then tracing out the third qubit.

\begin{figure}[h]
	\begin{center}
		\includegraphics[angle=0, width=0.45\textwidth]{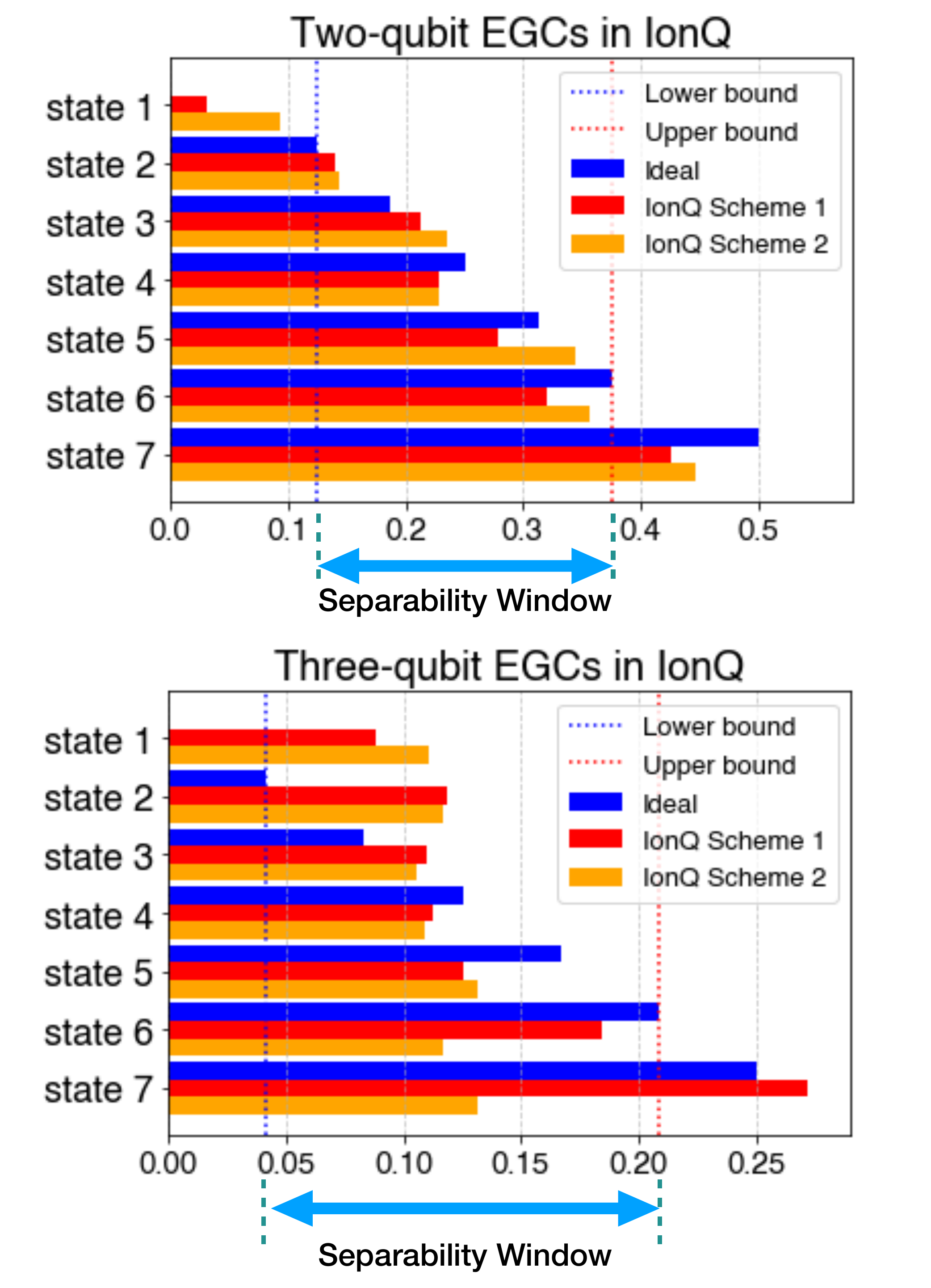}
		\caption{  An EWC is realized in IonQ to certify two- and three-qubit EGCs. For two-qubit EGCs, lower and upper bounds are $1/8=0.125$ and $3/8=0.375$, respectively, violations of which certify an EGC. That is, the separability window is $[0.125,0.375]$: values estimated out of the range certify entangled states. The violations are observed for state 1 and state 7, entangled ones in Equation \ref{eq:2state}. Both $Scheme~1$ and $Scheme~2$ show the certification. \\
			For three-qubit EGCs, the bounds are $1/24\approx 0.417$ and $5/24 \approx 0.2083$: the separability window is given by $[0.417, 0.2083]$. Violations of the upper bound are observed by $Scheme~1$ for state 7 in Equation \ref{eq:3state}. These are the first results that entanglement generation is certified in IonQ where the command of qubit allocations is lacking.}  
\label{2qubitionq}
	\end{center}	
\end{figure}

\section{ Demonstration in IonQ }
\label{sec:ionq}

IonQ can be accessed via Azure Quantum and Amazon Braket. The physical system applies identical ytterbium atoms prepared in an ultra-high vacuum chamber, where each ytterbium ion is identified as a single qubit. The ions are trapped by rapidly oscillating electromagnetic fields and then, logic gates are implemented by laser control on the atoms. Since qubits can move under a control, the system has an advantage in the connectivity. Gate fidelities are relatively high, over $99\%$ for single-qubit gates and $98\%$ for two-qubit gates on average, which may allows a longer-depth circuit. 

\subsection{2 qubit EGCs}
The EWC in Equation \ref{eq:4ewpure} is realized in IonQ in both $Scheme~1$ and $Scheme~2$. The demonstration of certifying EGCs has been performed on Jan. 13 and 14 2021. The result is shown in Figure \ref{2qubitionq}. We recall that IonQ does not provide the command of qubit allocations. Hence, one can find that the results certifies entanglement generation without qubit allocations.

\subsection{3 qubit EGCs}
\label{subsec:3qubit}

We have performed the certification of three-qubit EGCs in IonQ on Jan. 14 2021. Both $Scheme~1$ and $Scheme~2$ are exploited to realize the EWC. Since the EW 2.0 operator in Equation \ref{eq:spa3ew} is implemented by three qubits, additional three qubits are needed to construct its purification, 
\bea
    \ket{\widetilde{W}}_{1\cdots3^{'}}  &=& U_{\widetilde{W}} |0\rangle_{1231^{'}2^{'}3^{'}}^{\otimes 6}\nonumber\\
    &=&\sqrt{\frac{1}{8}} \left( \sqrt{\frac{2}{3}}  \ket{\phi^-_{000}} \otimes \ket{010} + \sqrt{\frac{2}{3}} \ket{\phi^+_{001}} \otimes \ket{111}   \right. \nonumber \\
    & & \left. + \sqrt{\frac{4}{3}} \ket{\phi^-_{101}} \otimes \ket{101} + \sqrt{\frac{4}{3}} \ket{\phi^+_{010}} \otimes \ket{001}  \right. \nonumber \\
    & & \left. + \sqrt{2} \ket{\phi^-_{010}} \otimes \ket{100} + \sqrt{\frac{2}{3}} \ket{\phi^+_{011}} \otimes \ket{110} \right. \nonumber \\
    & & \left. + \sqrt{\frac{4}{3}} \ket{\phi^-_{011}} \otimes \ket{011} \right). ~~~~~\label{eq:uw3}
\eea
We recall that $\ket{\phi^{\pm}_{\x}} =  (\ket{\x} \pm \ket{\bar{\x}})/\sqrt{2}$ for a sequence $\mathrm{x} = \x_1 \x_2 \x_3$ and $\bar{\x} = \bar{\x}_1 \bar{\x}_2 \bar{\x}_3$ with $\bar{\x}_1 =  1-\x_1$, e.g, $\ket{\phi^+_{000}} =  (\ket{000} + \ket{111}) /\sqrt{2}$. From Equation \ref{eq:ew3}, the lower and the upper bounds are $1/24 \approx 0.0417$ and $5/24 \approx 0.2083$, respectively. We also recall that the entangled state $|\phi_{000}^+ \rangle$ is detected by the lower bound, and the state $ |\phi_{010}^-\rangle$ by the upper bound. Since EWs are robust to noise, entangled states close to those states can be also detected. 

The certification circuit $U_{\widetilde{W}}$ is shown in Figure \ref{3circuit}. Note that the single-qubit gates in the circuits are given to perform the following transformations, 
\bea
R_Y (\theta_1) : |0\rangle & \mapsto & \sqrt { \frac{2}{3}} |0\rangle + \sqrt { \frac{1}{3}} |1\rangle \nonumber \\
R_Y (\theta_2) :  |0\rangle  & \mapsto & \sqrt { \frac{1}{3}} |0\rangle + \sqrt { \frac{2}{3}} |1\rangle \nonumber \\
R_Y (\theta_3) |0\rangle & \mapsto &  \sqrt { \frac{3}{5}} |0\rangle + \sqrt { \frac{2}{5}} |1\rangle \nonumber \\
R_Y (\psi_1) : |0\rangle & \mapsto &  \sqrt { \frac{1}{4}} |0\rangle + \sqrt { \frac{3}{4}} |1\rangle  \nonumber \\
R_Y (\psi_2) : |0\rangle & \mapsto & \sqrt { \frac{3}{8}} |0\rangle + \sqrt { \frac{5}{8}} |1\rangle \nonumber\\
R_Y (\xi) : |0\rangle & \mapsto & \sqrt { \frac{1}{3}} |0\rangle + \sqrt { \frac{2}{3}} |1\rangle  \label{eq:3gate}
\eea

Then, quantum circuits generating the following states are considered in the proof-of-principle demonstration: \\
\bea
\mathrm{  state ~1}: &~~  &   |\phi_{000}^+\rangle =  (|000 \rangle + |111 \rangle)/\sqrt{2} \nonumber \\
\mathrm{  state ~2}: & ~~&  |000\rangle \langle 000|   \nonumber \\
\mathrm{  state ~3}: & ~~&  (1/2)|000\rangle \langle 000| + (1/2)|001\rangle \langle 001|  \nonumber \\
\mathrm{  state ~4}: & ~~&  |001\rangle \langle 001|  \nonumber \\
\mathrm{  state ~5}:  & ~~ &  (1/2)|001 \rangle \langle 001| + (1/2)|010 \rangle \langle 010 |  \nonumber \\
\mathrm{  state ~6}:  & ~~ &  |010 \rangle \langle 010|  \nonumber \\
\mathrm{  state ~7}: & ~~&   |\phi_{010}^-\rangle =  (|010\rangle - |101\rangle)/\sqrt{2} . \label{eq:3state}
\eea
Note that states 1 and 7 are entangled and the others are separable. When realizing mixed states, state $3$ and state $5$, in a circuit, one can prepare their purifications with one ancila qubit. For instance, the state $3$ is obtained by preparing the following state, 
\bea
|\phi_3\rangle = \frac{1}{\sqrt{2}} |001\rangle |0\rangle +  \frac{1}{\sqrt{2}} |110\rangle |1 \rangle \nonumber
\eea
and then tracing out the fourth one. The circuits to prepare state $3$ and state $5$ are realized by using four qubits.

\section{Discussion}
\label{sec:discussion}

In the certification, a measurement is repeated $8192$ times in the case of IBMQ and $2000$ times in IonQ. The standard deviation is about $0.001$, which thus does not affect the conclusion in the certification. Note that preparations and measurements are assumed to be independently and identically distributed.

As for the certification of three-qubit EGCs where the circuit depth also increases, it is observed that IonQ is more reliable: the certification has been shown in IonQ, where the circuit depth is about $130$. Note that $Scheme~2$ has been also considered since it contains an advantage of having a short-depth circuit, which is comparable to that of a preparation circuit. It, at the same time, requires more resources in preparation of qubits, the number of two-qubits gates in the projection step, and more measurements. These allow a higher chance of having errors in state preparation and two-qubit gates, all of which are significant in NISQ technologies. 

\subsection{Estimation of entanglement generation}

The result presented can be incorporated to quantifying entanglement generation by PQCs in NISQ algorithms such as QAOA or VQE, where shallow circuits are repeatedly applied. In general, expectation values of EWCs give a lower bound to entanglement measures. To this end, let us introduce an entanglement measure for two-qubit EGC,
\bea
E[U_G]  = \max_{|\psi\rangle_{12}} E[U_G |\psi\rangle_{12}] \nonumber
\eea
where $E[|\psi\rangle_{12}]$ denotes the entropy of entanglement for the state $|\psi\rangle_{12}$ \cite{PhysRevA.54.3824}, i.e., $E[|\psi\rangle_{12}] = S(\tr_i |\psi\rangle_{12}\langle \psi|)$ for $i=1,2$ where $S$ denotes the von Neumann entropy. 

From the results in Refs. \cite{PhysRevA.72.022310, PhysRevLett.98.110502, PhysRevA.75.052302}, the entanglement measure above is bounded lower as follows,
\bea
E[U_G] \geq -\min ( \tr[W\rho],0 ). \nonumber
\eea
This shows that entanglement generation by an unknown circuit $U_G$ can be quantified by observing EWs or equivalently EWCs. Since EW 2.0 contains two bounds in terms of $p$ and $q$ in Equation \ref{eq:bound}, one can derive the followings 
\bea
E[U_G] &\geq&  \frac{-1}{1-p}  \min [  (p_{U_{\widetilde{W}}} (0^{\times 2n}) [U_G]  - \frac{p}{4}, 0 ],  \nonumber\label{eq:measure1} \\
\mathrm{and}  ~~E[U_G] &\geq&  \frac{-1}{1-q}  \min [  p_{U_{\widetilde{W}}} (0^{\times 2n}) [U_G]  - \frac{q}{4},  0 ].   \nonumber \label{eq:measure2}
\eea 
Applying the certification data in Figures \ref{2qubitibmq2} and \ref{2qubitionq}, one can in fact quantify entanglement generation in the cloud quantum computing services:
 \begin{center}
\begin{tabular}{l*{6}{c}r}
     Entanglement  && IBMQ & IonQ  \\ \hline
    \hline
E [$U_G$ for state 1] & \vline&   0.1728  &   0.1890 (Scheme 1)  \nonumber \\ 
E  [$U_G$ for state 7]& \vline & 0.0896 &    0.1420 (Scheme 2) \nonumber  \\ 
\end{tabular}
\end{center}
Therefore, we have shown that estimation of entanglement generation can also be obtained from the certification of EGCs.

\section{ Conclusion}
\label{sec:conclusion}

A cloud-based quantum computing service does not precisely correspond to a typical scenario of a quantum laboratory. The qubit allocation in a cloud computing service is not in the users' hands but by the service, whereas it is by experimenters in a quantum laboratory. It is an additional assumption to trust qubit allocations by a cloud-based computing service. Two classical data are available to a user of the service,  firstly a design of a quantum circuit as an input, and secondly, the statistics of measurement outcomes as an output.

We have established the framework of certifying entanglement generation in a cloud-based quantum computing service without the assumption of trusting qubit allocations. Namely, the design of quantum circuits is presented by which entanglement generation is certified by interpreting measurement outcomes returned by a cloud service. The entanglement certification circuits are constructed by transferring the framework of EW 2.0 to a quantum circuit model via SPA to EWs. Henceforth, the certification of entanglement generation in cloud-based quantum computing would be valid no matter how unsuccessful qubit allocations are in a cloud computing service.

We have used the circuit architecture to certify entanglement generation in IBMQ and IonQ services. In particular, entanglement generation in two and three qubits by CNOT and Toffoli gates is certified: namely, two- and three-qubit entangled states are certified. Since they form a universal set of gates together with single-qubit gates, entanglement generation in an arbitrary $n$-qubit circuit may be certified in a similar manner. We have also provided a lower bound to the capability of entanglement generation from the experimental data obtained in the certification.

Our results can be generally used to certify entanglement generation in a cloud quantum computing service. Quantum hardware in which entanglement generation is certified meets an essential requirement to gain quantum advantages. We envisage that when practical information tasks are attempted in cloud-based quantum computing services, the proposed framework applies to find the capability of entanglement generation toward achieving quantum advantages. In future investigations, it would be interesting to find if more of the assumptions apart from qubit allocations may be relaxed so that the level of the certification becomes even higher. For instance, it may be possible to certify semi-device-independent quantum correlations that may be generated in cloud quantum computing, e.g. \cite{VanHimbeeck2017semidevice}.


\section*{Acknowledgement}
This work is supported by the National Research Foundation of Korea (NRF-2021R1A2C2006309 and NRF-2019M3E4A1080001) and Samsung Research Funding \& Incubation Center of Samsung Electronics (Project No. SRFC-TF2003-01).


%

\end{document}